\documentclass[letters,usedcolumn,a4paper,fleqn,usenatbib]{mnras}

\usepackage{txfonts}

\usepackage[T1]{fontenc}
\usepackage{ae,aecompl}

\usepackage{graphicx}	
\usepackage{amssymb}	

\newcommand{\xmm}{{\it XMM-Newton}}
\newcolumntype{d}[1]{D{.}{.}{#1}}

\title[XUV irradiation of Earth-sized planets in TRAPPIST-1]{Strong XUV irradiation of the Earth-sized exoplanets orbiting the ultracool dwarf TRAPPIST-1}

\author[P.\ J.\ Wheatley et al.]{
Peter J. Wheatley,$^{1}$\thanks{E-mail: P.J.Wheatley@warwick.ac.uk}
Tom Louden,$^{1}$
Vincent Bourrier,$^{2}$
David Ehrenreich$^{2}$\newauthor
and Micha\"el Gillon$^{3}$
\\
$^{1}$Dept.\ of Physics, University of Warwick, Gibbet Hill Road, Coventry CV4 7AL, UK\\
$^{2}$Observatoire de l'Universit\'e de Gen\`eve, 51 chemin des Maillettes, 1290 Versoix, Switzerland\\
$^{3}$Institut d'Astrophysique et de G\'eophysique, Universit\'e de Li\`ege, All\'ee du 6 Ao\^ut 19C, 4000 Li\`ege, Belgium
}

\date{Accepted 2016 September 16. Received 2016 August 22.}

\pubyear{2016}

\begin{document}
\label{firstpage}
\pagerange{\pageref{firstpage}--\pageref{lastpage}}
\maketitle

\begin{abstract}
We present an \xmm\ X-ray observation of TRAPPIST-1, which is an ultracool dwarf star recently discovered to host three transiting and temperate Earth-sized planets. We find the star is a relatively strong and variable coronal X-ray source with an X-ray luminosity similar to that of the quiet Sun, despite its much lower bolometric luminosity.
We find $L_{\rm X}/L_{\rm bol}=2-4\times10^{-4}$, with the total XUV emission in the range $L_{\rm XUV}/L_{\rm bol}=6-9\times10^{-4}$, 
and XUV irradiation of the planets that is many times stronger than experienced by the present-day Earth.
Using a simple energy-limited model we show that the relatively close-in Earth-sized planets, which span the classical habitable zone of the star, are subject to sufficient X-ray and EUV irradiation to significantly alter their primary and any secondary atmospheres. Understanding whether this high-energy irradiation makes the planets more or less habitable is a complex question, but our measured fluxes will be an important input to the necessary models of atmospheric evolution. 
\end{abstract}




\section{Introduction}
\citet{t1} announced the discovery of a remarkable system of three Earth-sized planets orbiting a nearby ultracool dwarf star of spectral type M8, TRAPPIST-1. The planets are transiting, providing precise radii, and because the host star is small and cool the transits are deep and the planets are temperate despite their relatively short orbital periods. 

The three planets are most likely all outside the classical habitable zone, two closer-in and one beyond (although the outer planet has an uncertain orbit that could place it in the habitable zone). Nevertheless, the factors that influence habitability are complex and uncertain 
and \citet{t1} point out that habitable conditions might exist at the terminators of the inner planets that are presumably tidally-locked, while tidal heating of the outer planet might render it habitable as well. Either way, the small size and low temperature of the star, and the proximity of the system to Earth (12\,pc), provide by far the best opportunity to date to study the atmospheres of cool, Earth-sized exoplanets. 

An important factor influencing the evolution of planetary atmospheres and their habitability is the X-ray 
 (1--124\,\AA)
and extreme-ultraviolet (EUV; 
 124--912\,\AA) radiation emitted 
by their parent stars (together often termed XUV radiation).
Mass loss from exoplanetary atmospheres is observed directly in ultraviolet transit observations of 
hot gas giants and a warm Neptune
\citep[e.g.][]{Vidal-Madjar03,Lecavelier12,Ehrenreich15} and this is thought to be the result of XUV irradiation heating of the planetary exosphere and driving hydrodynamic escape \citep[][]{Lammer03,Johnstone15}. The long term effects of XUV irradiation on the habitability of terrestrial planets are complex and uncertain, and while some planets might be rendered uninhabitable through atmospheric stripping, others may become habitable through the removal of a massive primary atmosphere of H/He \citep[e.g.][]{Owen16}. Water might be removed from some habitable zone planets by photolysis and H evaporation \citep[e.g.][]{Bolmont17}, perhaps leading to abiotic oxygen-dominated atmospheres \citep{Wordsworth14,Luger15}, while in other planets the evaporation might prevent the atmosphere of an out-gassing planet from becoming too dense. It has also been suggested that XUV irradiation 
might expand a secondary atmosphere beyond the magnetosphere of the planet, where it becomes vulnerable to erosion by the stellar wind \citep[e.g.][]{Lammer11}.

Ultracool dwarfs are known to exhibit stellar activity, but the activity level seems to decrease steeply to later spectral types, with $L_{\rm X}/L_{\rm bol}$ values dropping by at least two orders of magnitude from saturated emission of $10^{-3}$ for mid-M stars \citep[e.g.][]{Pizzolato03, Wright11} to $<10^{-5}$ for mid-L dwarfs \citep[e.g.][]{Berger10}. \citet{Williams14} confirm the breakdown of saturated X-ray emission for spectral types later than M6, but find that a population of objects later than M7 with X-ray emission characteristic of mid-M stars is not excluded. 

TRAPPIST-1 has been found to exhibit chromospheric $H\alpha$ emission at a level of $L_{H\alpha}/L_{\rm bol}=2.5-4.0\times10^{-5}$, which is found to be typical for its M8 spectral type and weaker than seen in mid-M stars \citep{Gizis00,Reiners10}. TRAPPIST-1 also has a relative weak magnetic field strength of $\sim 600\,\rm G$ \citep{Reiners10}, which is lower than mid-M stars with the same short spin period of $1.40\pm0.05$\,d. It may therefore be expected to have X-ray emission considerably weaker than mid-M stars, and indeed \citet{Bolmont17} assumed $L_{\rm X}/L_{\rm bol}<10^{-5}$ in a recent study of water loss from the Earth-sized planets of TRAPPIST-1. 

In this letter we present an \xmm\ observation of TRAPPIST-1 that allows us to measure the X-ray luminosity of the star, estimate its EUV luminosity, and hence consider the effects of XUV irradiation on the Earth-sized exoplanets. 

\section{Observations}
The host star of the TRAPPIST-1 system (=2MASS\, J23062928--0502285) was observed with \xmm\ for 30\,ks on 17th December 2014 (ObsID: 0743900401; PI: Stelzer) using the thin optical blocking filters. An X-ray source is clearly detected at the proper-motion corrected 2MASS position of the ultracool dwarf \citep{Cutri03,Costa06}. The source is soft, being visible in pipeline processed EPIC-pn images in the 0.2-0.5 and 0.5-1.0\,keV bands, but not in the higher energy bands. The \xmm\ pipeline source detection also identifies a source at this position, with a offset of 3.1\,arcsec from the expected source position. This offset is consistent 
with the known accuracy of the \xmm\ astrometric frame \citep{Watson09} and we find the offset drops to 1.27\,arcsec when the \xmm\ astrometry is rectified against the USNO B1.0 catalogue. 

We extracted X-ray lightcurves and spectra for TRAPPIST-1 from the EPIC-pn camera using the pipeline source detect position and a 20\,arcsec radius aperture. The EPIC-pn camera observed for 28.0\,ks and had an effective exposure time of 24.9\,ks. For such a soft source only a small proportion of the X-ray events are detected in the EPIC-MOS cameras so we limited our analysis to the EPIC-pn. The background counts were estimated using a source-free circular region of radius 51.5\,arcsec located at the same end of the same CCD. We followed the standard data reduction methods as described in data analysis threads provided with the Science Analysis System\footnote{http://www.cosmos.esa.int/web/xmm-newton/sas} (SAS version 14.0). The spectrum was binned to a minimum of 10 counts per bin, with the additional requirement that the EPIC-pn spectral resolution would not be oversampled by more than a factor of 3. We fitted the spectrum using XSPEC\footnote{https://heasarc.gsfc.nasa.gov/xanadu/xspec/} (version 12.8). Our fitted parameters were determined using the Cash statistic \citep{Cash79} and our quoted errors correspond to 68\% confidence intervals.

A number of background flares occurred during the observation (caused by soft protons from the Sun impacting the detector), however we did not filter these time intervals when creating the data products presented here (as suggested in the SAS threads) because we wanted to inspect the entire X-ray light curve and measure X-ray fluxes averaged across the observation (see Sect.\,\ref{sec-res}). 

\section{Results}
\label{sec-res}
\subsection{X-ray light curve}
\label{sec-lc}
\begin{figure}
\begin{center}
	\includegraphics[width=7.5cm]{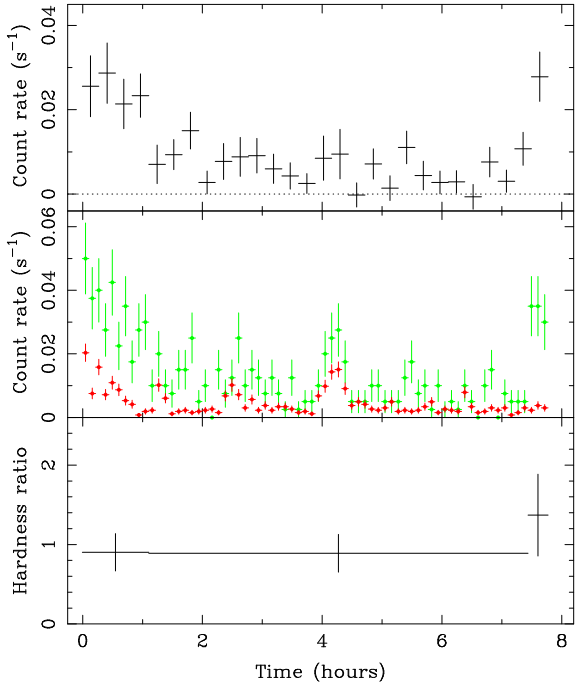}
\end{center}
    \caption{ Top: the X-ray light curve of TRAPPIST-1 (0.15--1.5\,keV) with the \xmm\ EPIC-pn camera (1000\,s bins). 
Middle: the raw source light curve of TRAPPIST-1 (green) and the scaled background (red), showing the Solar proton flares (400\,s bins). 
Bottom: the hardness of TRAPPIST-1 calculated as the ratio of X-ray counts in the 0.5--1.5\,keV and 0.15--0.5\,keV bands.}
    \label{fig-lc}
\end{figure}

The \xmm\ X-ray light curve of TRAPPIST-1 is plotted in Fig.\,\ref{fig-lc}, showing that the star was brighter at the beginning and the end of the observation. This variability is statistically significant, with a $\chi^2$ of 85.6 for 27 degrees of freedom when compared to the weighted mean of all the data points. The 7.8\,hr observation covers 23\% of the 1.40\,d spin period of the star \citep{t1} so it is likely that at least some of this variability is due to rotational modulation.  

For completeness we also plot the raw source and background time-series in  
Fig.\,\ref{fig-lc} (middle panel), showing the effect of the Solar soft proton flares. 
While both TRAPPIST-1 and the background were bright at the beginning of the observation, we are satisfied that this is a coincidence.  
Tests with light curves extracted from higher energies (where TRAPPIST-1 is not detected) confirm that the background subtraction is robust.

In the bottom panel of Fig.\,\ref{fig-lc} we plot a measure of the hardness of the X-ray spectrum of TRAPPIST-1 during the beginning, middle and end of the observation. Hardness ratios are a simple method with which to identify variations in the X-ray spectrum, and in this case we have calculated the ratio of the X-ray count rate in the 0.5--1.5\,keV band to the count rate in the 0.15--0.5\,keV band. 
To the precision of the current dataset, it can be seen that the hardness ratio is consistent with the X-ray spectrum remaining constant throughout the observation, despite the flux variations apparent in the light curve.  There is a hint that the hardness may have increased during the brightening at the end of the observation, and an increases in hardness would be expected if this brightening were due to a stellar flare. 

\subsection{Spectral analysis}
\label{sec-spec}
\begin{figure}
\begin{center}
	\includegraphics[width=7.5cm,height=6.5cm]{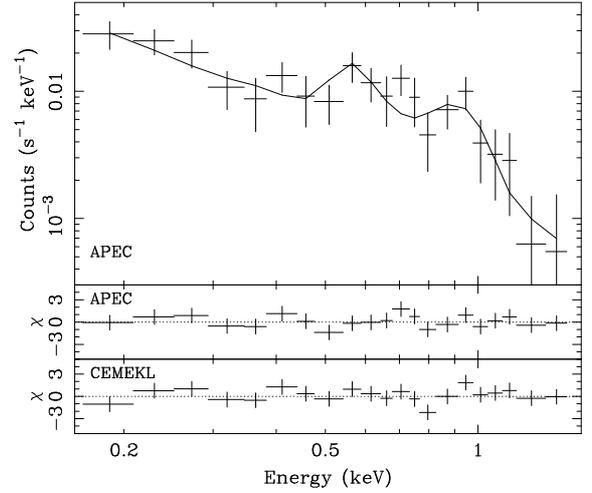}
\end{center}
    \caption{The \xmm\ EPIC-pn spectrum of TRAPPIST-1 fitted with our two temperature APEC model (top). The middle panel shows the normalised residuals to this fit, while the bottom panel shows the residuals to our fit with the {\it cemekl} model. }
    \label{fig-spec}
\end{figure}

The \xmm\ EPIC-pn spectrum of TRAPPIST-1 is plotted in Fig.\,\ref{fig-spec}. The X-ray spectrum is very soft and shows evidence of line emission between 0.5 and 1.0\,keV. These features are characteristic of 
coronal emission from late type stars. We fitted the spectrum using the APEC model for a collisionally-ionised optically-thin plasma \citep{Smith01}, finding a poor fit with a single temperature model, but a good fit with a two temperature model ($\chi^2$ of 11.9 with 17 degrees of freedom). The model and residuals to this fit are plotted in the top and middle panels of Fig.\,\ref{fig-spec} respectively. The fitted temperatures are $kT=0.15\pm^{0.02}_{0.01}$ and $0.83\pm^{0.16}_{0.10}$\,keV. In reality, coronal X-ray emission is expected to be the sum of emission from a wide and continuous range of temperatures \citep[e.g.][]{Louden17}, but at low spectral resolution and with modest signal-to-noise ratios a two temperature model usually provides an adequate approximation \citep[e.g.][]{Pillitteri14}. 

As expected for such a nearby X-ray source \citep[$12.1\pm0.4$\,pc;][]{Costa06} we found that the interstellar X-ray absorption is too low to be constrained usefully by the X-ray spectrum. Consequently we chose to fix the interstellar absorption in our models at a value of $N_{\rm H}=3.7\times10^{18}\,\rm cm^{-2}$ based on an assumed local interstellar neutral hydrogen density of $0.10\,\rm cm^{-3}$ \citep{Redfield00}. We modeled the absorption with the {\it tbabs} model in XSPEC \citep{Wilms00}, and found that it had a negligible effect on our fitted temperatures and X-ray fluxes. 

Elemental abundances are also poorly constrained by the X-ray spectrum and we left them fixed at Solar values \citep{Asplund09}.

Our fitted X-ray energy fluxes for TRAPPIST-1 are presented in Table\,\ref{tab-flux}, together with X-ray luminosities calculated from the known distance to the star \citep[$12.1\pm0.4$\,pc;][]{Costa06}. We have given values for a range of energy intervals in order to facilitate comparison with other studies of the coronal X-ray emission of late type stars. 

In order to calculate our X-ray fluxes in the commonly-used {\it ROSAT} PSPC band (0.1--2.4\,keV) it was necessary to extrapolate our fitted model beyond the soft cut-off of our EPIC-pn X-ray spectrum at 0.166\,keV. This extrapolation is sensitive to the number and distribution of temperature components employed in the model, and it is possible that our simple two temperature model under-predicts the X-ray flux in the 0.1--2.4\,keV band. In order to investigate the uncertainty in this extrapolation we also fitted our spectrum with the {\it cemekl} model in XSPEC,  which calculates the X-ray spectrum of an optical-thin plasma with a continuous range of temperatures (up to a maximum value) and with the emission measure distribution defined by a power law 
\citep[][]{Schmitt90,Singh96}. 
We used the version of {\it cemekl} based on the same APEC model employed in our two temperature fit. We find an almost equally good fit to the spectrum with this model ($\chi^2$ of 15.7 with 19 degrees of freedom) and the residuals are plotted in the lower panel of Fig.\,\ref{fig-spec}. In this fit the emission measure rises steeply to lower temperatures, with a power law index of $-0.81\pm0.15$. The maximum temperature is poorly constrained to $kT_{\rm max}>1.23$\,keV and we left this parameter fixed to $kT_{\rm max}=5$\,keV while evaluating the X-ray fluxes. 

The fluxes and corresponding luminosities for the {\it cemekl} fit are also presented in Table\,\ref{tab-flux}, and it can be seen that the fluxes of the two-temperature APEC model and the {\it cemekl} model are consistent within the energy band covered by the EPIC-pn spectrum, but diverge as the model is extrapolated down to 0.1\,keV. This is as expected because the two temperature model does not account for the emission from cooler plasma that may contribute significantly below the EPIC-pn band, while the power law model for the distribution of emission measures may be too steep below the EPIC-pn band and over-predict the contribution from lower temperatures. In Sect.\,\ref{sec-dis} we assume that the true 0.1-2.4\,keV X-ray luminosity falls between the best fitting values from these model extremes, i.e. in the range $(3.8-7.9)\times10^{26}\,\rm erg\,s^{-1}$. 

\section{Discussion and Conclusions}
\label{sec-dis}
Our spectral analysis in Sect.\,\ref{sec-spec} and X-ray luminosities in Table\,\ref{tab-flux} show that TRAPPIST-1 is a relatively strong coronal X-ray source. It has the same 0.1--2.4\,keV X-ray luminosity as the quiet Sun \citep[$6\times10^{26}\,\rm erg\,s^{-1}$,][]{Judge03} despite its photospheric luminosity of only $0.000525\pm0.000036\,\rm L_{\odot}$ \citep{Filippazzo15}. 

The $L_{\rm x}/L_{bol}$ ratio of the star is in the range $(2-4)\times10^{-4}$, which places it below the canonical value of $10^{-3}$ for saturated X-ray emission of stars with spectral types G to mid-M \citep[e.g.][]{Pizzolato03,Wright11}. TRAPPIST-1 is a reasonably rapidly rotating star, with a spin period of $1.40\pm0.05\,\rm d$, and so it might be expected to exhibit saturated X-ray emission for its spectral type. The relatively low flux compared to the canonical saturated value might then reflect the known decrease in stellar activity to spectral types later than M6 \citep[e.g.][]{Berger10,Williams14}.  However, inspection of the distribution of $L_{\rm x}/L_{bol}$ values for individual earlier-type saturated stars in \citet{Wright11} shows a considerable spread around the mean value of $7\times10^{-4}$, and many earlier type stars have $L_{\rm x}/L_{bol}$ in the range $(2-4)\times10^{-4}$ that we observe for TRAPPIST-1. Consequently, the X-ray emission of TRAPPIST-1 can also be considered to be consistent with the saturated emission of earlier type M stars. 

In order to consider the possible effects of X-ray and EUV irradiation on the atmospheres and possible oceans of the Earth-sized planets orbiting TRAPPIST-1 we estimate energy-limited mass loss rates \citep[e.g.][]{Lammer03,Lecavelier07,Louden17} as
\begin{equation}
\dot{M} = \frac{\eta {\rm \pi} F_{XUV} \alpha^2 R_{P}^3}{G M_p K} = \frac{\eta I_{XUV} \alpha^2 R_{P}}{G M_p K}
\end{equation}
where $F_{XUV}$ is the combined X-ray and EUV fluxes incident on the planet, $I_{XUV}$ is the total X-ray and EUV irradiation of the planet, G is the gravitational constant and $M_p$ and $R_p$ are the mass and radius of the planet respectively. The factor $K$ accounts for the reduced energy required to escape the Roche lobe of the planet \citep{Erkaev07}. 
We set the quantity $\alpha$ to unity, which is designed to take account of the increased cross-sectional area of planets to EUV radiation. This is an important correction for hot gaseous planets, but probably negligible for cooler terrestrial planets. 
$\eta$ is the energetic efficiency of mass loss, which has been constrained observationally to be at least 1\% in hot gas giants and warm Neptunes \citep[e.g.][]{Lecavelier12,Ehrenreich15} and is expected to be around 10--20\% for low mass planets \citep[e.g.][]{Owen16a}.

In order to calculate energy-limited escape rates we need to estimate the EUV flux of the star, which is not covered by the \xmm\ bandpass. To do this we employ the scaling relation of \citet{Chadney15}, which is an empirical relationship between the X-ray flux at the surface of the star and the relative strength of the X-ray and EUV emission. We cannot be sure that this relation applies to such a late spectral type as TRAPPIST-1, but we are encouraged that the \citet{Chadney15} study includes a mid-M star, and that our measured surface X-ray flux for TRAPPIST-1 ($4.6-9.5\times10^{5}\rm \,mW\,m^{-2}$) lies in the middle of the range calibrated by the empirical relation. Using this relation we find $F_{\rm EUV}/F_{\rm X}=1.78$ for the surface flux calculated from our APEC spectral fit (Table\,\ref{tab-flux}) and $F_{\rm EUV}/F_{\rm X}=1.31$ for our {\it cemekl} model. 
While our EUV fluxes are estimated by a different method to that employed by the MUSCLES Treasury Survey (where they are scaled from the reconstructed Lyman-$\alpha$ fluxes), we note that the two methods have been found to yield consistent estimates for the M dwarf planet host GJ\,436 \citep{Ehrenreich15,Youngblood16}.

Summing 
X-ray and EUV fluxes we find $L_{\rm XUV}/L_{\rm bol}=6-9\times10^{-4}$, which is almost an order of magnitude higher than any of the planet hosts in the MUSCLES Treasury Survey \citep{France16}. In Table\,\ref{tab-escape} we show how our fluxes translate to irradiation of the individual planets ($F_{\rm X}$, $F_{\rm EUV}$ and $I_{\rm XUV}$), together with the energy-limited mass loss rates for each planet (assuming an energetic efficiency of $\eta=0.1$ and taking fluxes from the APEC model; the {\it cemekl} values being simply a factor of 1.7 higher). 
It can be seen that the high-energy irradiation of the planets is typically tens to over a thousand times higher than experienced by the present-day Earth, giving likely escape rates that would be highly significant for Earth-like planets with atmospheric masses of around $5\times10^{21}\,\rm g$ and ocean masses of around $1\times10^{24}\,\rm g$. On the timescale of a Gyr, all three planets could have have been stripped of atmospheres and oceans. Even at the wider possible separations, TRAPPIST-1d could be very substantially eroded, including for instance the entire H component of the UV photo-dissociated water content of the Earth. 
\begin{table}
	\begin{center}
	\caption{Fitted X-ray fluxes and luminosities for TRAPPIST-1 in different energy bands. APEC refers to our two temperature model. {\it cemekl} is  our multi-temperature model where the emission measure distribution is defined by a power law.}
	\label{tab-flux}
	\begin{tabular}{ccccc} 
Energy range& \multicolumn{2}{c}{X-ray flux$\rm^a$} & \multicolumn{2}{c}{Luminosity$\rm^b$}\\
(keV) & APEC & {\it cemekl} & APEC & {\it cemekl}\\\hline
0.100 -- 2.40 & $2.16\pm^{0.18}_{0.21}$ & $4.49\pm^{0.44}_{0.70}$ 
& $3.79\pm^{0.36}_{0.41}$ & $7.89\pm^{0.86}_{1.28}$\\
0.124 -- 2.48 & $2.06\pm^{0.15}_{0.18}$ & $2.94\pm^{0.19}_{0.36}$ 
& $3.62\pm^{0.31}_{0.36}$ & $5.16\pm^{0.41}_{0.68}$\\
0.150 -- 2.40 & $1.98\pm^{0.13}_{0.19}$ & $2.42\pm^{0.13}_{0.31}$ 
& $3.48\pm^{0.28}_{0.37}$ & $4.25\pm^{0.30}_{0.58}$\\
0.200 -- 2.40 & $1.83\pm^{0.11}_{0.16}$ & $1.88\pm^{0.09}_{0.22}$ 
& $3.21\pm^{0.24}_{0.32}$ & $3.30\pm^{0.22}_{0.42}$\\
	\end{tabular}
        \end{center}
$\rm^a\ \times 10^{-14}\,\rm erg\,s^{-1}\,cm^{-2}$ \\
$\rm^b\ \times 10^{26}\,\rm erg\,s^{-1}$ 
\end{table}

\begin{table*}
	\begin{center}
	\caption{The X-ray and EUV irradiation of the individual Earth-sized planets in the TRAPPIST-1 system. The symbols $F_{\rm x}$ and $F_{\rm EUV}$ denote the energy fluxes at the planet, while $I_{\rm XUV}$ is the total X-ray and EUV energy input to each planet. Mass loss rates were calculated assuming energy-limited atmospheric escape with an efficiency of 10\%. 
The figures here are based on our best-fitting APEC fluxes of Table\,\ref{tab-flux} and it should be noted that our {\it cemekl} fits suggest that all of these figures could be higher by a factor 1.7. Orbital separations and planet radii are from \citet{t1}.}
	\label{tab-escape}
	\begin{tabular}{clcd{4.1}d{4.1}d{4.1}d{4.1}d{2.3}d{2.3}d{2.2}} 
Planet & Separation & Radius & \multicolumn{1}{c}{$F_{\rm X}$} & \multicolumn{1}{c}{$F_{\rm EUV}$} & \multicolumn{1}{c}{$F_{\rm X}$} & \multicolumn{1}{c}{$F_{\rm EUV}$}& \multicolumn{1}{c}{$I_{\rm XUV}$} & \multicolumn{2}{c}{Estimated mass loss} \\
name & (AU) & ($\rm R_{Earth}$) & \multicolumn{2}{c}{($\rm erg\,s^{-1}\,cm^{-2}$)}& \multicolumn{2}{c}{(cf. Earth$\rm^c$)} & \multicolumn{1}{c}{($\times10^{20}\,\rm erg\,s^{-1}$)} & \multicolumn{1}{c}{($\rm \times 10^7\, g/s$)} & \multicolumn{1}{c}{(Earth oceans$\rm^d$/Gyr)} \\\hline
TRAPPIST-1b & 0.01111   & 1.113& 1092.& 1950.& 1280.& 644. &48.1 & 118. & 29. \\
TRAPPIST-1c & 0.01522   & 1.049& 582. & 1039.&  682.& 343. &22.7 & 47.2 & 12. \\
TRAPPIST-1d & 0.022$^a$ & 1.168& 278. & 497. &  326.& 164. &13.5 & 29.6 & 7.2 \\
TRAPPIST-1d & 0.058$^b$ & 1.168& 40.1 & 71.6 &  47.0& 23.7 &1.94 & 3.85 & 0.93\\
TRAPPIST-1d & 0.146$^a$ & 1.168&  6.3 & 11.3 &   7.4&  3.7 &0.31& 0.59& 0.14\\
	\end{tabular}
        \end{center}
$\rm^a$ The minimum and maximum possible orbital separations for TRAPPIST-1d. \hfill
$\rm^b$ The most likely orbital separation for TRAPPIST-1d.\\
$\rm^c$ The Earth typically receives $0.85\,\rm erg\,s^{-1}\,cm^{-2}$ in X-ray and $3.03\,\rm erg\,s^{-1}\,cm^{-2}$ in EUV in mid-Solar cycle \citep{Ribas05}.\\
$\rm^d$ Taken to be $1.3\times10^{24}\rm\,g$.
\end{table*}

On the other hand, energy-limited mass loss is rather simplistic and can only provide an upper limit to mass loss rates, neglecting as it does the radiation physics and hydrodynamics of the planetary atmosphere and its composition. \citet{Owen16} for instance show that energy limited mass loss models can considerably over-estimate escape rates. They also show that rather strong XUV irradiation is actually required for a terrestrial planet to become habitable if it is formed with a substantial H/He primary atmosphere. 

\citet{Bolmont17} have carried out an investigation of the likely rates of water loss from Earth-sized exoplanets in the habitable zones of ultracool dwarfs, and in TRAPPIST-1 in particular. They conclude that TRAPPIST-1b and -1c are likely  to be completely desiccated by XUV irradiation, but that TRAPPIST-1d may have held onto most of its initial water content. However, these authors assume $L_{\rm XUV}/L_{\rm bol}<10^{-5}$ for TRAPPIST-1, which is at least fifty times smaller than the value we measure here. On the face of it this seems to make a significant water content on TRAPPIST-1d also unlikely, although \citet{Bolmont17} do list a number of mechanisms that influence water loss and require further investigation. Water might, for instance, survive in cold traps on the night sides or at the poles of highly-irradiated tidally-locked planets \citep[e.g.][]{Leconte13,Menou13}.

The TRAPPIST-1 system presents a fabulous opportunity to study the atmospheres of Earth-sized planets as well as the complex and uncertain mechanisms controlling planet habitability. 
Whatever the mechanisms at play, 
it is clear that these planets are subject to X-ray and EUV irradiation that is many-times higher than experienced by the present-day Earth and that is sufficient to significantly alter their primary and any secondary atmospheres. The high-energy fluxes presented here are vital inputs to atmospheric studies of the TRAPPIST-1 planets.

\section*{Acknowledgements}
P.W. is supported by a Science and Technology Facilities Council (STFC) consolidated grant (ST/L000733/1). T.L. is supported by a STFC studentship. V.B and D.E. are supported by the Swiss National Science Foundation (SNSF) through the National Centre for Competence in Research ``PlanetS''. Based on observations obtained with XMM-Newton, an ESA science mission with instruments and contributions directly funded by ESA Member States and NASA.

\bibliographystyle{mnras}
\bibliography{paper} 
\bsp	
\label{lastpage}
\end{document}